\documentclass[12pt]{article}

\ifx\pdfoutput\undefined
   \usepackage[dvips]{graphicx}
   \else
   \usepackage[pdftex]{graphicx}
   \fi
   \usepackage{epstopdf}

\usepackage{times}

\def\lta{\lower2pt\hbox{$\buildrel {\scriptstyle <}
   \over {\scriptstyle\sim}$}}
\def\gta{\lower2pt\hbox{$\buildrel {\scriptstyle >}
   \over {\scriptstyle\sim}$}}

\def\tfb{t_{\rm fb}}
\def\tacc{t_{\rm acc}}

\title{Properties of Gamma-ray Burst Progenitor Stars}

\author{Pawan Kumar$^1$, Ramesh Narayan$^2$ \& Jarrett L. Johnson$^1$ \\
$^1$Astronomy Department, University of Texas, Austin, TX 78712\\
$^2$Harvard-Smithsonian Center for Astrophysics, Cambridge, MA 02138}

\date{}

\begin{document}

\DeclareGraphicsRule{.tif}{png}{.png}{`convert #1 `basename #1 .tif`.png}

\baselineskip24pt

\maketitle 

\begin{quote}
{\bf We determine some basic properties of stars that produce spectacular 
gamma-ray bursts at the end of their life. We assume that accretion of the 
outer portion of the stellar core by a central black hole fuels the prompt 
emission, and that fall-back and accretion of the stellar envelope later
produces the plateau in the x-ray light curve seen in some bursts.
Using x-ray data for three bursts we estimate the radius 
of the stellar core to be $\sim(1-3)\times10^{10}\,$cm, and 
that of the stellar envelope to be $\sim(1-2)\times10^{11}\,$cm. The 
density profile in the envelope is fairly shallow, with $\rho \sim r^{-2}$.
The rotation speeds of the core and envelope are $\sim0.05$
and $\sim0.2$ of the local Keplerian speed, respectively.  }
\end{quote}


\begin{figure}
\begin{center}
\includegraphics[width=5.95in]{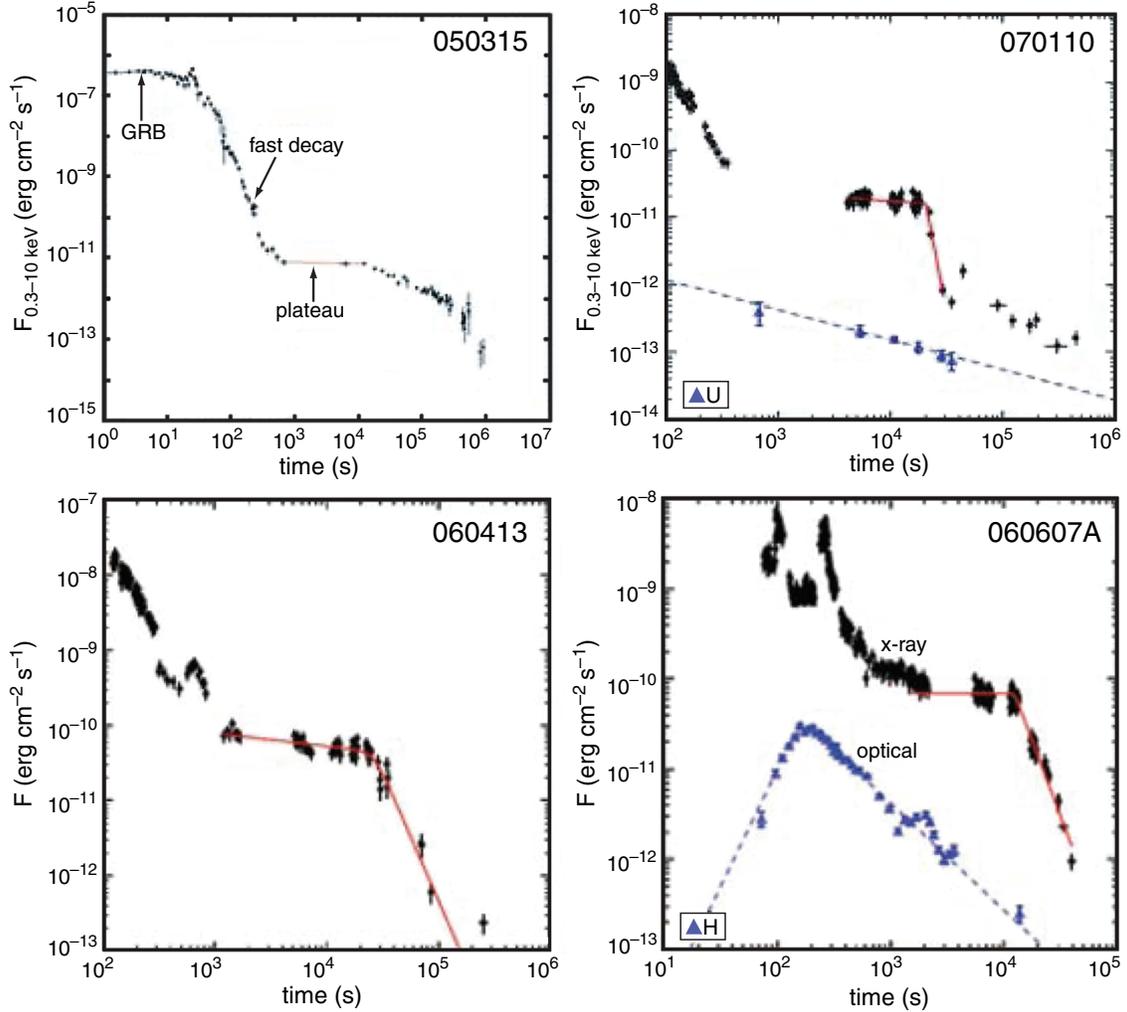}
\caption{\small The top left panel (from Vaughan et al. 2006) shows the 
0.3-10 keV band x-ray lightcurve (LC) of a typical long duration burst, 
GRB 050315. Four distinct phases are seen in the LC: the prompt GRB 
lasting for about a minute, an early steep decline lasting for $\sim 10$ 
minutes, a plateau lasting until about $10^4\,$s, and a ``normal'' post-plateau
decay. The top right panel (from Liang et al. 2007) shows x-ray and optical
LCs for GRB 070110.  The x-ray LC falls very sharply at the end of the
plateau ($t\sim 3\times10^3\,$s), before switching to a normal
decline, whereas the optical LC is a single power-law for the entire
duration. The two lower panels (also from Liang et al. 2007) show the
x-ray LCs of GRB 060413 and GRB 060607A.  
Note the complex x-ray LCs of these bursts, with flares, breaks, and plateaus.
In contrast, the optical LCs of GRBs are typically smooth and simple, such 
as those of GRB 070110 and GRB 060607A, shown in the right panels in blue. }
\label{fig1}
\end{center}
\end{figure}

\renewcommand{\thefootnote}{\fnsymbol{footnote}}

Observations of gamma-ray bursts (GRBs) suggest that the activity at
the center of these explosions lasts for several hours (1,2). The most
compelling evidence is provided by three bursts (3)  --
GRBs 060413, 060607A and 070110 -- which show a sudden decline in
their x-ray light curves (LCs) a few hours after the prompt burst
(Fig. 1).  The flux decline is by a factor of 10 or more and is much too
sharp for the radiation to originate in an external forward shock (4);
 the most likely explanation is continued
activity at the center of the explosion until at least the time of the
decline.
Additional evidence for continued activity of the central engine is
provided by the x-ray flares seen in many GRBs (5-7), and also by those
bursts whose x-ray \& optical afterglow lightcurves (LCs) are mutually
incompatible with a common origin (8, 9). In fact, central engine activity is 
implicated whenever the observed flux variability time scale, $\delta t$, 
is much smaller than the time elapsed since the onset of explosion.
The reason is that for a relativistic external shock causality dictates 
that $\delta t \gta R/2c\Gamma^2$, and the time it takes for photons to
arrive at the observer from a shock front at radius $R$ that is moving
with Lorentz factor $\Gamma$ is also $\sim R/2c\Gamma^2$, i.e. 
$\delta t/t \sim 1$ for external shocks. 

In this paper, we adopt the collapsar model of GRBs (10,11), in which the 
inner part of the progenitor star
collapses to a rapidly-spinning black hole (BH) and the remaining gas
from the star accretes on to the BH and produces an ultra-relativistic
jet. With a few plausible assumptions, we show that x-ray
observations of the three bursts mentioned above (GRBs 060413, 060607A
\& 070110) may be inverted to infer the structure and rotation
rate of their progenitor stars. These particular bursts were selected 
for this work because they have prominent
x-ray plateaus which are almost certainly the result of 
central engine activity. Thus for these bursts, we have information on
the power generation at the center covering an extended period of time,
which enables us to determine the core and envelope structure of the
progenitor star; only three bursts met this strict requirement of central
engine activity dominated x-ray plateau. Some of the results we find regarding 
progenitor star properties -- especially the core structure -- apply to 
a much larger sample of bursts, as discussed at the end of the paper. 
We also note that
optical emission was detected for two of the three bursts (Fig. 1). The 
optical LC is consistent with origin in the forward shock (FS).
Moreover, for a reasonable set of parameters the amount of x-ray flux 
produced in the FS is found to be $\sim10^2$ times smaller than the 
flux observed during the plateau.  Thus, there is consistency between 
a FS origin for the optical LC and central engine activity for the x-ray
emission.


The panel on the left in Fig. 2 shows a schematic GRB x-ray lightcurve,
which is based on the data shown in Fig. 1.
The panel on the right outlines the basic features of our model
and shows how the four phases of the LC are
connected to corresponding zones in the progenitor star.

We assume that the star has a core-envelope structure, as is common in 
stellar models. The bulk of the mass is in the stellar core.  

Some of the mass in the core collapses directly to form a spinning
BH and the rest accretes on the BH to produce the prompt GRB emission.
Surrounding the inner zone is a second zone, which represents the
transition region between the core and the envelope.  This zone has a
steeply falling density profile, and there is correspondingly a rapid
decline in the x-ray flux.  The outermost zone is a relatively
low-density stellar envelope.  Accretion of this gas produces the
plateau in the LC. The drop in the x-ray flux at the end of the plateau, 
which can be sudden, corresponds to the time when the outermost
layers of the envelope are accreted.

\begin{figure}[h!]
\begin{center}
\includegraphics[width=6.6in]{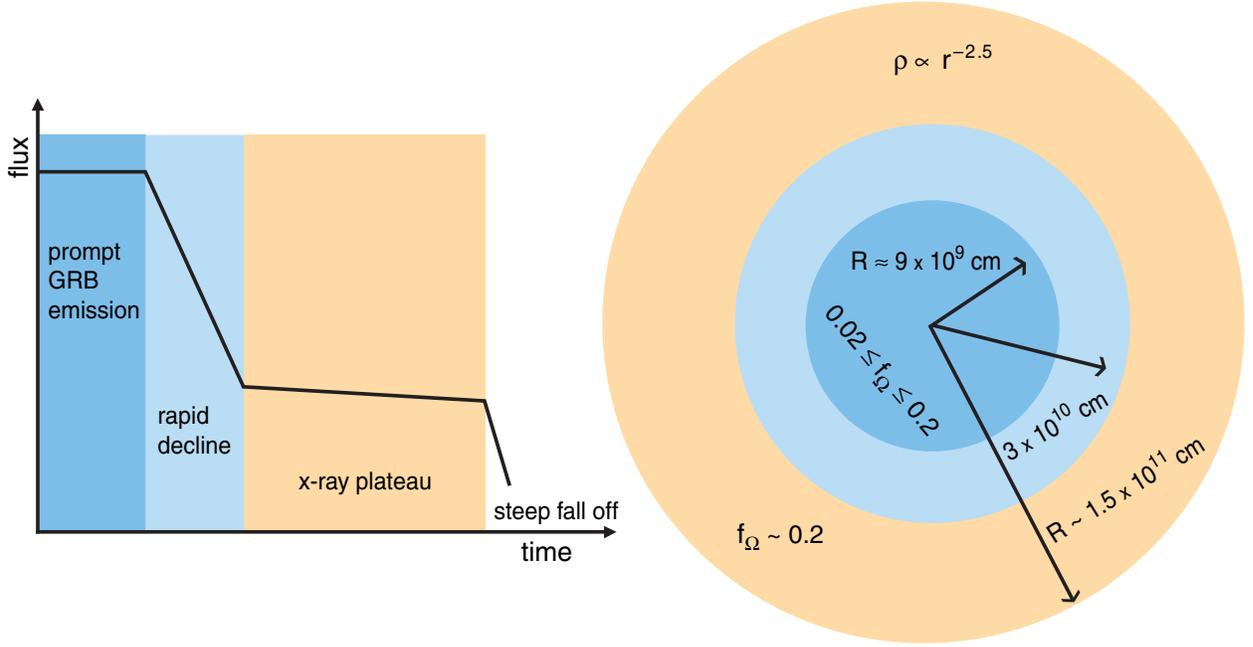}
\caption{\small The panel on the left shows a schematic x-ray lightcurve
with the following four segments: a prompt emission phase, a steep decline
phase, a plateau phase, and a post-plateau phase.  For the three GRBs
considered in this work, the last phase has a steep and sudden
decline. The panel on the right outlines our proposal for how the different 
segments in the LC are related to the accretion of corresponding zones in 
the progenitor star. The radii ($r$) and spin parameters ($f_\Omega\equiv 
\Omega/\Omega_k$) of the various zones are estimated from the x-ray data. }
\label{fig2}
\end{center}
\end{figure}


The fall-back time is the time it takes a parcel of gas
in the progenitor star at radius $r$ to fall to the center, and it is
approximately equal to the free fall time (12),
 $\tfb\sim 2(r^3/GM_r)^{1/2}$, where $M_r$ is the mass
interior to $r$, and all times are in the frame of the GRB host
galaxy.  Let us assume that the accretion time $\tacc$ of the gas in
the disk is smaller than $\tfb$ (see below).  Then, at a given time
$t=10^2t_2\,$s, the central engine will accrete gas that has fallen
back from a radius $r=10^{10}r_{10}\,$cm, where
\begin{equation}
  r_{10}\sim 1.5 t_2^{2/3} M_{\rm BH,1}^{1/3}\, ,
\label{rfb}
\end{equation}
and $M_{\rm BH,1}$ is the BH mass in units of 10$M_\odot$.  We are
assuming here that the interior mass $M_r$ is dominated by the BH.
Let us suppose that the gas at radius $r$ in the progenitor star has
an angular velocity equal to a fraction $f_\Omega$ of the local
Keplerian velocity:
\begin{equation}
\Omega(r) = f_\Omega(r)\Omega_k(r) = f_\Omega(r)\, (GM_r/r^3)^{1/2}.
\label{fOmega}
\end{equation}  
Then, the specific angular momentum $j_*=10^{18} j_{*,18} \,{\rm
cm^2\,s^{-1}}$ of this gas is given by
\begin{equation}
j_{*,18}(r) \approx 3.8\, M_{\rm BH,1}^{1/2} r_{10}^{1/2} 
 f_\Omega(r) \approx 4.7 
   \, t_{2}^{1/3} M_{\rm BH,1}^{2/3} f_\Omega(r).
\label{jstar}
\end{equation}

The gas will fall back toward the black hole until its angular momentum 
allows it to become centrifugally supported, at which point it becomes
part of a thick accretion disk about the black hole.
Its further in fall will then occur on the time scale at
which viscous forces in the disk allow the orbit to decay.

The viscous accretion time of the fall-back gas after it has
circularized is approximately equal to $\tacc\sim
2/\alpha\Omega_k(r_d)$, where $\alpha=0.1\alpha_{-1}$ is the viscosity
parameter, and $r_d\sim j_*^2/G M_{\rm BH}$ is the radius at which gas
with specific angular momentum $j_*$ circularizes.  Here we have
assumed that the accretion disk is geometrically thick, which is
generally true at the high mass accretion rates expected in a
collapsar (13).  We thus have
\begin{equation}
\tacc\sim 10\,\alpha_{-1}^{-1} f_\Omega^3(r) \tfb \sim 10\,
\alpha_{-1}^{-1} j_{d,18}^3M_{\rm BH,1}^{-2} \,{\rm s}\,.
\label{tacc}
\end{equation}
The first relation shows that the accretion time will be shorter than
the fall-back time so long as $f_\Omega\ \lta\ 0.4$ (for
$\alpha\sim0.1$).

 The jet power from the central engine is determined by the
mass accretion rate $\dot{M}_{\rm BH}$ on to the BH, which is usually a
small fraction of the mass fall-back rate $\dot{M}_{\rm fb}$ on to the
accretion disk (the majority of the fall-back mass is expelled in a 
disk wind as it spirals in toward the BH [14]).  

If the mass fall-back rate
decreases suddenly at some time $t_0$, e.g., because of the transition
from the stellar core to the envelope or when gas at the outer edge of
the envelope has fallen back, then it can be shown by the
conservation of mass and angular momentum that the accretion rate on
the BH will decline with time as (15)
\begin{equation}
\dot{M}_{\rm BH}(t) \sim \dot{M}_{\rm BH}(t_0)\left[1 +
1.5(t-t_0)/\tacc\right]^{-2}, \quad t \ge t_0.
\label{mdot}
\end{equation}
The exponent outside the square brackets is model-dependent and varies
between $-4/3$ and $-8/3$, depending on details; for definiteness, we
choose $-2$.  If $\tacc \ll t_0$, the effect of the above time
dependence is that, in a $\log$-$\log$ plot, the jet power will
initially drop by a large factor $\sim(t_0/\tacc)^2$ within a time
$\sim t_0$, and the power will then transition to an asymptotic
decline $\propto t^{-2}$.  This feature in the predicted LC plays an
important role in our model.  It is how we explain the sudden decline
in the x-ray LC at the end of the prompt GRB and at the end
of the plateau.

We apply these scaling relations to derive constraints on the
stellar structure and rotation rate of GRB progenitors.
A number of the features seen in GRBs can be explained naturally within 
the collapsar model as corresponding to the accretion of different portions 
of the progenitor star onto the central BH. Here we quantify the size, 
density, and rotation rate of the stellar core and envelope, as inferred 
from the observed timescales discussed above.


Observations with the x-ray telescope aboard the Swift satellite show
that the flux in the 0.3--10 keV band of a typical long duration GRB
is roughly constant for about $20\,$s (Fig. 1, top left panel), after
which the x-ray LC undergoes a rapid decline as $\sim t^{-3}$ or
faster for about 5 minutes.
Assuming that it takes $\sim30\,$s for the BH to form, and adding to
this the burst duration of $20\,$s, we estimate that the end of the
main burst is approximately $50\,$s after the initiation of core
collapse.  By equation (\ref{rfb}), the gas that falls back at this
time comes from a radius $\sim 9\times10^{9}M_{\rm BH,1}^{1/3}\,$cm.  This
radius must correspond to the outer edge of the innermost zone in the
schematic stellar model shown in Fig. 2.


The subsequent steeply declining phase after the initial burst goes
from $t\sim50-300\,$s.  In our model, this portion of the LC is
associated with accretion of gas from the second zone in the star, the
transition region between the core and the envelope where the density
has a steep decline.  Using the same argument as before, the outer
radius of the second zone must be $\sim 3\times10^{10}M_{\rm
BH,1}^{1/3}\,$cm.

A rapid decline in the x-ray LC requires a rapid decline in
$\dot{M}_{\rm BH}$ (faster than $\sim t^{-2}$), which is possible only
when the disk is able to adjust quickly to a decrease in the
mass-fall-back rate $\dot{M}_{\rm fb}$.  For an accretion flow to
respond quickly to rapid changes in $\dot{M}_{\rm fb}$ at time $t$ we
must have $\tacc\ll t$ (see Eq. \ref{mdot}).  This condition provides
an upper limit on the rotation speed in the core region of the
progenitor star.  The x-ray flux typically drops by a factor of $\sim
10^3$ during the rapid decline phase at the end of the prompt burst
(Fig. 1). Therefore, by Eq. \ref{mdot}, we require
$(250/\tacc)^2\ \gta\ 10^3$ or $\tacc\ \lta\ 8\,$s at $t\sim50\,$s.
 From Eq. 4, this means we must have $f_\Omega\ \lta\
0.2\alpha_{-1}^{1/3}$ at $r\approx 10^{10}\,$cm.

We can obtain a lower limit on $f_\Omega$ in the stellar core by
requiring that the angular momentum should be sufficiently large that
the fall-back gas is able to form an accretion disk, i.e., $r_d\ \gta\
3 R_g$, where $R_g\equiv GM_{\rm BH}/c^2$ (note that the radius of the
innermost stable circular orbit is $2.3R_g$ for a BH with a spin
parameter of 0.9).  Since during core collapse a particle initially at
$r$ ends up at a radius $r_d\approx r[f_\Omega(r)]^2$, where it is
centrifugally supported, the condition $r_d\ \gta\ 3 R_g$ yields
$f_\Omega(r)\ \gta\ 0.02$. An additional constraint on the core
rotation rate is provided by the total energy of $\sim 10^{52}\,$erg
observed in a typical GRB. Numerical simulations show that the
efficiency for jet production from a rapidly rotating BH is about 1\%
(16), and therefore the total mass accreted by the BH
should be about $0.5 M_\odot$. This suggests that much of the mass
within $r\approx 10^{10}\,$cm collapses directly to the BH, and
therefore $f_\Omega(r)$ is not much larger than 0.02; a reasonable
estimate is $f_\Omega \sim 0.05$ for $r\ \lta\ 10^{10}\,$cm.

\medskip

A plateau in the x-ray LC is seen typically from $\sim 3\times10^2 -
3\times10^3\,$s (time in host galaxy rest frame). Gas falling on the
central accretion disk during this time interval arrives from $3\times
10^{10} M_{\rm BH,1}^{1/3}\,{\rm cm}\ \lta\ r\ \lta\ 1.5\times 10^{11}
M_{\rm BH,1}^{1/3}\,$cm (Eq. \ref{rfb}).  We suggest that the gas
accreted during this phase comes from the envelope of the progenitor
star, which means that the outer radius of the envelope must be $\sim
1.5\times 10^{11}$~cm.  In principle, the plateau could be produced by
supernova ejecta that failed to escape.  However, such a model is
unlikely to give the sharp cutoff which is seen in the x-ray LC at the
end of the plateau in the three GRBs considered in this paper. For
these GRBs at least, the plateau must arise from fall-back
of the stellar envelope, which has a well-defined outer edge.

The density profile in the stellar envelope can be determined from the
rate of decline of the x-ray flux during the plateau: $f_x\propto
t^{-\delta}$, $\delta\sim0.5$. If the specific angular momentum of the
gas in the envelope is constant, as might be the case if the envelope
is convective, then the required density profile is $\rho(r)\propto
r^{-3(\delta+1)/2} \sim r^{-2.2}$ for $\delta\sim0.5$ (15).
 For a convective stellar envelope that is not very
massive, we expect $\rho \propto r^{-3/2}$ when gas pressure dominates
and $\rho\propto r^{-3}$ when radiation pressure dominates. The
density profile we infer, assuming a constant specific angular
momentum envelope, lies between these two limits. If the specific
angular momentum increases with increasing radius, then a shallower
decline of density is indicated; for instance, $\rho\propto r^{-1.5}$
if the specific angular momentum increases with distance as $\sim
r^{3/4}$. A solid-body rotation profile for the envelope is ruled out
because it requires an unphysical density structure in which $\rho$ is
almost independent of $r$.

\smallskip

The x-ray flux at the end of the plateau in GRBs 060413, 060607A \&
070110 falls off sharply by a factor $\sim10^2$. The fall-off occurs
at $t\sim3\times10^3\,$s and extends over a time $\delta t\sim t$.
This steep fall-off requires the accretion time to be sufficiently
short, $\tacc\ \lta\ t/10\sim 300\,$s, which implies that $j_{d,18}\
\lta\ 3\alpha_{-1}^{1/3} M_{\rm BH,1}^{2/3}$ and $f_\Omega(r)\ \lta\
0.2 \alpha_{-1}^{1/3}$ at $r\sim 1.5\times 10^{11}$ (17, 18).

A lower limit for $f_\Omega$ in the stellar envelope is obtained by
the requirement that the in-falling gas should be able to form an
accretion disk at the center; this implies $f_\Omega\ \gta\ 0.01$.
Another constraint comes from the fact that, apart from the flares
discussed below, the x-ray LC during the plateau is usually quite
smooth. This suggests that $\tacc$ is fairly large (which plays the role of
a smoothing time scale), and so $f_\Omega$ is probably not much less
than the upper limit of $\sim0.2$.

After the sharp fall-off at the end of the plateau, the LC is expected
to decline as $\sim t^{-2}$ if the x-ray emission is dominated by jet
luminosity associated with the debris disk (see Eq. \ref{mdot}).
However, at these low flux levels the observed x-rays might be
dominated by emission from shock-heated circumburst gas, i.e. external
shock emission. In this case the flux will decline as $\sim t^{-1}$, e.g., 
GRB 070110 (Fig. 1).

For those GRBs where the x-ray LC makes a smooth transition at the end
of the plateau phase, $f_\Omega$ should be such that $\tacc\sim t$.
For these systems, we require $f_\Omega\sim 0.4$ near the outer edge
of the progenitor star.


Rapid flares with short rise times are often seen during the plateau
phase of the x-ray LC, and these provide additional constraints.  A
flare was seen at the beginning of the x-ray plateau in GRBs 060413
and 060607A, and at the end of the plateau in GRB 070110 (Fig. 1), 
with a rise time of order $0.1t$.  
Assuming the flares are produced by a disk instability, the rise time
should be a factor of a few larger than the instability time scale 
$t_{\rm inst}$. For a viscous instability $t_{\rm inst} \sim \tacc$ 
whereas for a  dynamical (e.g., gravitational) instability 
$t_{\rm inst}\sim \Omega_k^{-1} \sim\alpha\,\tacc$. Taking the flare rise time 
to be $\sim 5 t_{\rm inst}\sim 0.1 t$ we find that 
$f_\Omega\sim 0.1\alpha_{-1}^{1/3}$ 
in the stellar envelope if flares arise as a result of viscous instability, 
and $f_\Omega\sim 0.3$ for a dynamical instability origin for flares. 
The amplitude of the flare $f_{x,{\rm
flare}} /f_{x,{\rm plateau}}$ can be at most $\sim\tacc/t_{\rm inst}$. 
Thus, even in the limit of a dynamical instability, the amplitude is
limited to $\sim\alpha^{-1}\sim10$. Much larger flare amplitudes
(e.g., in GRB 060526 [5]) might suggest an unusually small value of the
viscosity parameter $\alpha$.  Alternatively, these flares may be
caused by a sudden increase in the mass fall-back rate, though such an
event is not easy to visualize in our model.

The decay time of flares is expected to be of order $\tacc$, as this is the 
time scale on which a transient enhancement of the accretion rate, regardless 
of its origin, will subside. A noticeable difference
between the x-ray plateau and the prompt emission is that LCs are typically
more variable during the burst, and this raises a question as to why the
central engine behaves differently during the two phases. Part of the
difference is that instability timescales are longer when the outer envelope 
of the star, with a larger specific angular momentum, is accreted. Another 
factor is that, during the plateau phase, the jet propagates through an 
already evacuated cavity and is less prone to fluctuating baryon loading.


An upper limit on the stellar radius $R_*$ is provided by the
requirement that the energy expended as the relativistic jet makes its
way out of the star not exceed the energy initially injected into the
jet.  Using results in (19, 20)
we find $R_*\ \lta\ 5\times10^{11}\,$cm.  Our radius estimates are
consistent with this limit.

A possible way to get a handle on the mass of the GRB progenitor star
is via the total energy produced in the explosion.  The mass accreted
during the prompt burst was estimated in \S2.2.2 to be $\sim0.5M_\odot$.
The energy release during the plateau is typically about 10\% of that
in the prompt burst, i.e., $\sim 10^{51}\,$erg (assuming the same
beaming factor and efficiency as during the initial burst). Therefore,
the mass accreted by the BH is only $\sim 0.03 M_\odot$.  
Even after allowing for the fact that only a small fraction
$\sim(r_d/R_g)^{1/2}\sim10^{-2}$ of the total fall-back mass actually
reaches the BH, the rest being carried away in a disk wind, we estimate
that the mass of fall-back gas in the plateau phase is no more than a
few $M_\odot$. The mass of the BH can be constrained because
much of the mass within $\sim10^{10}\,$cm collapses to the BH; for an
evolved star this mass is about $5M_\odot$.  Thus we estimate the mass
of the GRB progenitor star to be $M_*\sim10M_\odot$, plus whatever
mass is ejected in the accompanying supernova explosion, which we are
unable to constrain.


Figure 2 summarizes our primary results on the properties of GRB
progenitor stars.  The three bursts we considered in this paper
are notable in that they have a steep fall-off of x-ray flux at the 
end of the plateau phase.  These bursts provide the most detailed
information on the properties of their progenitor stars. Many of our 
arguments and results, especially those in which we use the prompt burst
and subsequent steep fall-off to infer the properties of the stellar core,
should apply to any long duration GRB.  Similarly,
our discussion of the x-ray plateau can be applied to other
bursts that have plateaus in their x-ray LCs which are produced 
by continuous accretion activity i.e., those GRBs that show a simple
power-law decline of the optical LC and a plateau in the x-ray; 
these comprise roughly one third of the long GRBs observed by Swift.

\section{References and Notes}

\begin{enumerate}

\item The possibility of a long lived central engine activity was 
suggested$^2$ soon after the discovery of GRB afterglow emission.

\item J.I. Katz, T. Piran, R. Sari, Phy. Rev. Letters, 80, 1580 (1998)

\item E-W Liang, B-B Zhang, B. Zhang arXiv:0705.1373  (2007)

\item E. Nakar, J. Granot, astro-ph/0606011  (2006)

\item  D. Burrows et al., Science, 309, 1833 (2005)
 
\item  G. Chincarini, astro-ph/0702371  (2007)

\item B. Zhang, ChJAA  7, 1 (2007)
 
\item Y. Fan, T. Piran, MNRAS 369, 197 (2006)

\item A. Panaitescu et al. MNRAS 369, 2059  (2006)
 
\item S.E. Woosley, ApJ 405, 273 (1993)
 
\item B. Paczynski, ApJ 494, L45 (1998)

\item The factor of 2 in the expression for $\tfb$ accounts for the time it 
takes signals to travel to $r$ and to communicate the loss of pressure 
support at the center.

\item R. Narayan, T. Piran and P. Kumar, ApJ 557, 949  (2001)

\item K. Kohri, R. Narayan and T. Piran, ApJ 629, 341  (2005)
 
\item P. Kumar, R. Narayan and J. L. Johnson, to appear in MNRAS (2008)
 
\item J.C. McKinney, ApJ 630, L5  (2005)

\item Note that the decline of the LC cannot be steeper than $\sim t^{-3}$; 
this is due to late arriving photons from outside the relativistic beaming
angle of $\Gamma^{-1}$ (18). Therefore, we can
only place an upper limit on $f_\Omega$ from the steepness of the
decline after the x-ray plateau. 
 
\item P. Kumar and A. Panaitescu, ApJ 541, L51 (2000)

\item E. Ramirez-Ruiz, A. Celotti, M.J. Rees,  MNRAS 337, 1349 (2002)

\item C.D. Matzner, MNRAS 345, 575 (2003)

\end{enumerate}

\end{document}